\documentclass[twocolumn,prl,aps,longbibliography,nobalancelastpage]{revtex4-2}
\usepackage{amsmath,amsfonts,amssymb,graphicx,hyperref,epstopdf,xcolor,tikz,scalerel,natbib,array,gensymb}
\usepackage{mathptmx,textcomp,braket}
\usepackage[T1]{fontenc}
\usepackage[utf8]{inputenc}
\usepackage{multirow}
\usepackage{array}
\usepackage{soul}
\DeclareMathAlphabet{\mathpcal}{OMS}{zplm}{m}{n}

\definecolor{lime}{HTML}{A6CE39}
\DeclareRobustCommand{\orcidicon}
{
	\begin{tikzpicture} 
	\draw[lime, fill=lime] (0,0) circle [radius=0.15] node[white] {{\fontfamily{qag}\selectfont \tiny ID}};
	\draw[white, fill=white] (-0.0625,0.095) 	circle [radius=0.007];
	\end{tikzpicture}
	\hspace{-2.2mm}
}

\newcommand\orcidID[1]{\href{https://orcid.org/#1}{\orcidicon}}

\newcommand{\be}{\begin {equation}}
\newcommand{\ee}{\end {equation}}
\newcommand{\beqa}{\begin {eqnarray}}
\newcommand{\eeqa}{\end {eqnarray}}

\newcommand{\Exp}[1]{\text{e}^{#1}}

\hypersetup{colorlinks,citecolor=blue,filecolor=black,linkcolor=blue,urlcolor=blue}
 
\begin{document}
\title{Harmonic emission as a probe to coherent transitions in the topological superconductors}
\author{Nivash R\orcidID{0009-0004-3076-9192}}
\email[E-mail: ]{nivash1807@gmail.com}
\author{S. Srinidhi\orcidID{0009-0004-3387-4908}}
\email{s.srinidhi312@gmail.com}
\thanks{\\ These authors contributed equally to this work.}
\author{Jayendra N. Bandyopadhyay\orcidID{0000-0002-0825-9370}}
\author{Amol R. Holkundkar\orcidID{0000-0003-3889-0910}}

\affiliation{Department of Physics, Birla Institute of Technology and Science Pilani, Pilani Campus, Vidya Vihar, Pilani, Rajasthan 333031, India.}

\begin{abstract}
 
 We investigate the dynamical behavior of a topological superconducting system, demonstrating that, its static configuration undergoes a transition driven by an intrinsic supercurrent. By analyzing the band population, we confirm the quasiparticle nature of the system both in the presence and absence of an external laser field. Under laser driving, we observe an enhancement in static emission forming a plateau-like structure, accompanied by multiple coherent transitions in the population. These transitions exhibit Rabi-like oscillations, attributed to the presence of Majorana bound states (MBS), further reinforcing the quasiparticle character of the model. Our results highlight the efficacy of laser driving as a probe of the system’s topological and dynamical stability.
\end{abstract}

\maketitle

The supercurrent is a dissipationless phenomenon that distinguishes superconductors from conventional conductors \cite{TSC1, TSC2}. In normal superconductors, Cooper pairs are formed through phonon-mediated interactions, as described by BCS theory. In the topological superconductors (TSCs), the pairing mechanisms are more complex, which include exchange of spin fluctuations, antiferromagnetic correlations, and other non-BCS interactions. These mechanisms of the pair formation leads to the emergence of topologically protected states \cite{Stewart_uncon_sc}. For example, the supercurrent in the Kitaev chain arises naturally due to the $p$-wave superconducting pairing, Majorana bound states (MBS) like topologically protected edge states, etc. \cite{Kitaev}. These properties of the topological superconductors made these materials a subject of extensive interest, particularly in exploring the behavior and stability of the supercurrent under various field conditions \cite{floquet1}, such as high-harmonic Generation (HHG) \cite{Corkum2007,RevModPhys,Ghimire2011,Ghimire2019}. Recent advances in HHG in solids \cite{Vampa2014,Luu2018,berry_18,Guan2020,Bai2021,Bionta2021,Neufeld_23,Nivash_AAH} involve the use of ultrafast and spatially structured fields with nanometric and attosecond precision \cite{Gouliemakis,Klemke2019,Li2020}, opening up exciting possibilities for novel coherent sources \cite{Takuya, Hohenleutner2015, Alexis_20, Recent_HHG_solids, Heide2024}. Coherent oscillations between quantum states arise when a system is driven by a resonant field, resulting in periodic population transfer between the states.
These oscillations are similar to Rabi oscillations \cite{Rabi_1937,Dudin2012}. This phenomenon is well-established in atomic and molecular systems, where these oscillations enable precise control over quantum state evolution and are central to the implementation of quantum logic operations \cite{Makhlin_2001,Chen2024}. The dynamics are typically governed by characteristic energy or momentum differences between the coupled states. In TSCs, the underlying complex pairing mechanisms enable coherent oscillations to emerge even in the absence of external driving. This behavior is surprising and significant because the system's intrinsic properties, particularly, its topologically protected states, like the MBS, naturally support oscillations between quantum states \cite{Nature_rabi, Rabi_Ramsey}. These oscillations are robust against disturbances because they are protected by the system’s topological nature, making them promising for fault-tolerant quantum computing \cite{Line_Defects_TSC}.

In this Letter, we focus on $p$-wave superconductors, where the Cooper pairs are formed due to the orbital motion rather than the spin interactions. Here we consider the spinless $p$-wave paired Kitaev chain \cite{Kitaev} as a prototype model of topological superconductors. We focus on how the combination of the supercurrent in the $p$-wave superconductor and the staggered hopping in the systems influence the population dynamics and gives rise to Rabi-like oscillations. The staggered hopping is controlled by a tuning parameter that introduces dimerization in the Kitaev chain. Recently, the dimerized Kitaev chain (DKC) has been studied extensively \cite{Sigrist_DKC,Shuchen_DKC, Roy2024_sreports_DKC, Shilpi_Roy_DKC}. Our work primarily focuses on the impact of external drives on these oscillations \cite{Liu2017,Ernotte,Nivash_23,Nivash_24}, offering insights into the stability of topologically protected states in the system. 

After introducing the $p$-wave SC Hamiltonian, we start investigation with the analysis of the static emission and harmonic spectra, through which the presence of an intrinsic supercurrent in the system is revealed. Later, we present an extensive study of the population dynamics, demonstrating how coherent transitions occur within the system, ultimately leading to Rabi-like oscillatory behavior \cite{Pop_Kitaev, Nature_Rabi_SC_qubit, Nature_rabi, Rabi_Ramsey}. The model Hamiltonian is of the form,
\begin{equation}
\begin{split}
H &=  \mu \sum_{j=1}^{N} \left( c_{j,A}^\dagger c_{j,A} + c_{j,B}^\dagger c_{j,B} \right) - \sum_{j=1}^{N-1} \Bigg\{ \Big[ w(1 - \lambda) c_{j,A}^\dagger c_{j,B} \\& + w(1 + \lambda) c_{j,B}^\dagger c_{j+1,A} \Big] -\Delta \Big[  c_{j,A}^\dagger c_{j,B}^\dagger + c_{j,B}^\dagger c_{j+1,A}^\dagger \Big] + \text{H.c.} \Bigg\},
\end{split}
\label{Ham}
\end{equation}
where $ c_{j,A}^\dagger $ (or $ c_{j,B}^\dagger $) is a fermionic creation operator on the sublattice A (or B) of the $ j $-th unit cell, $ w $ denotes the hopping potential, $ \mu $ is the chemical potential, and $ \Delta $ denotes the $p$-wave superconducting pairing potential, which is taken to be real in this case. The dimerization parameter $\lambda$ (where $ |\lambda| < 1 $) modulates the site-dependent hopping strengths. We have thoroughly analyzed the model elsewhere, where HHG is employed as a probe to explore the system's properties \cite{HHG_DKC}.
 
Here, we distinguish the bulk bands as follows: valence band (VB), and conduction bands as CB-I, CB-II, and CB-III; and the intermediate doubly degenerate states as MGS (mid-gap states) and MBS (Majorana-bound states), respectively \cite{Baldelli2022,Pattanayak2022}. Depending on the sign of their energy, we classify the midgap states as $+$ve midgap states (pMGS) and $-$ve midgap states (nMGS). Here in this Letter, our focus will be on the population dynamics of this system represented by Eq. \ref{Ham}. 

\begin{figure} 
  \includegraphics[width=1\columnwidth]{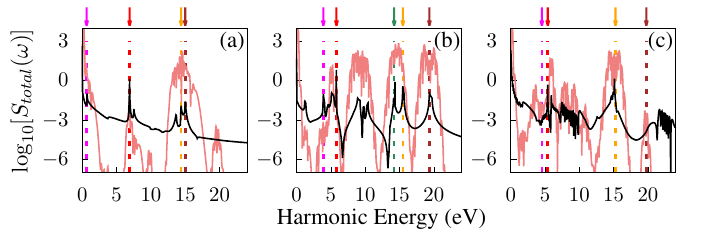}
   \caption{Figure depicts the results of the harmonic emission when the GS is chosen to be the conventional choice, i.e., the full VB and the first nMGS. The black line denotes the undriven GS when $A_0=0.0$, and the pink line corresponds to the driven case. Here $A_0 = 0.1$ a.u. The system parameters are $\mu = 0.225$ a.u., $w = 0.25$ a.u., $\Delta = 0.2375$ a.u. The tuning parameter $\lambda$ is kept to be (a) $\lambda=0.05$, (b) $\lambda=0.4$, and (c) $\lambda=0.8$. } 
   \label{fig2}
\end{figure}

Our investigation begins with the examination of the harmonic emission for different tuning parameters. In Fig. \ref{fig2} we have compared the HHG emission using the conventional GS ($\psi_{GS} =$VB + nMGS) with and without the driving field for different values of the dimerization parameter $\lambda$ \cite{Bauer2018,Bauer2019,Bian2022}. An external field is introduced in the system via the Peierl's phase by replacing the hopping term as $w \rightarrow w e^{-iaA(t)}$ in the field free Hamiltonian $H_0$, where $a$ is the lattice constant and $A(t)$ is the vector potential  \cite{Marlena2024,Chuan}. The vector potential is chosen to be a Gaussian pulse of amplitude $A_0=0.1$ a.u. with central wavelength to be $7.2 \mu$m and the full width at half maximum (FWHM) pulse duration to be $T=1.25\tau$ with the total time-duration $4T$. To compute the HHG spectrum, we calculate the total current operator,
\be
\begin{split}
 \mathfrak{J}(t)& =w (1-\lambda) \sum_{j=1}^{N-1} \left[ (-ia) \exp^{-iaA(t)} c_{j, A}^\dag c_{j, B} + \text{H.c.}\right] \\& + w(1+\lambda) \sum_{j=1}^{N-1} \left[(-ia) \Exp{-iaA(t)} c_{j, B}^\dag c_{j+1, A} + \text{H.c.}\right].
 \end{split}
 \ee
We solve the time-dependent Schr\"oendiger equation using the Crank-Nicholson method to obtain the time-evolved state of the system of the form $\ket{\psi (t+\delta t)} \sim e^{-iH(t)t} \ket{\psi_{GS}}$. The obtained HHG spectrum is presented in Fig. \ref{fig2}(a)-(c). The results corresponding to the static case ($A_0 = 0.0$) are presented as black lines, whereas the HHG spectra with $A_0 =0.1$ a.u. are shown using the pink lines. We see the sharp peaks for the static cases, arising from the intrinsic supercurrent in the TSC system. In Fig. \ref{fig2}(a), we observe the most significant peak around the red line region at $\lambda=0.05$, which corresponds to the transition between nMGS to MBS. 

We then increase the scale to $\lambda=0.4$, and we see multiple equiprobable emissions in the harmonic profile. But still, the order remains significant ($\sim 10^+1$) like the previous case, around the red line. The quasiparticle behavior remains unfolded in the emission spectra with transitions preserving the particle-hole symmetry in the system. In the previous case, we did not explain quasiparticle behavior. A new emission is seen for this choice of $\lambda$, which is marked using a green colored line, denoting the transition between VB and the last state of CB-II. As we further sweep into the scale, we set $\lambda=0.8$ and observe the harmonic profile. The most significant transition is observed near the orange line here. This denotes the transition between VB and pMGS. A lesser significant transition is also observed near the red line, implying nMGS $\leftrightarrow$ MBS. Interestingly, the harmonic profile exhibits a plateau-like structure beyond $5$ eV till $10$ eV, which was absent for the other cases.
	
\begin{figure*} 
  \includegraphics[width=1\textwidth]{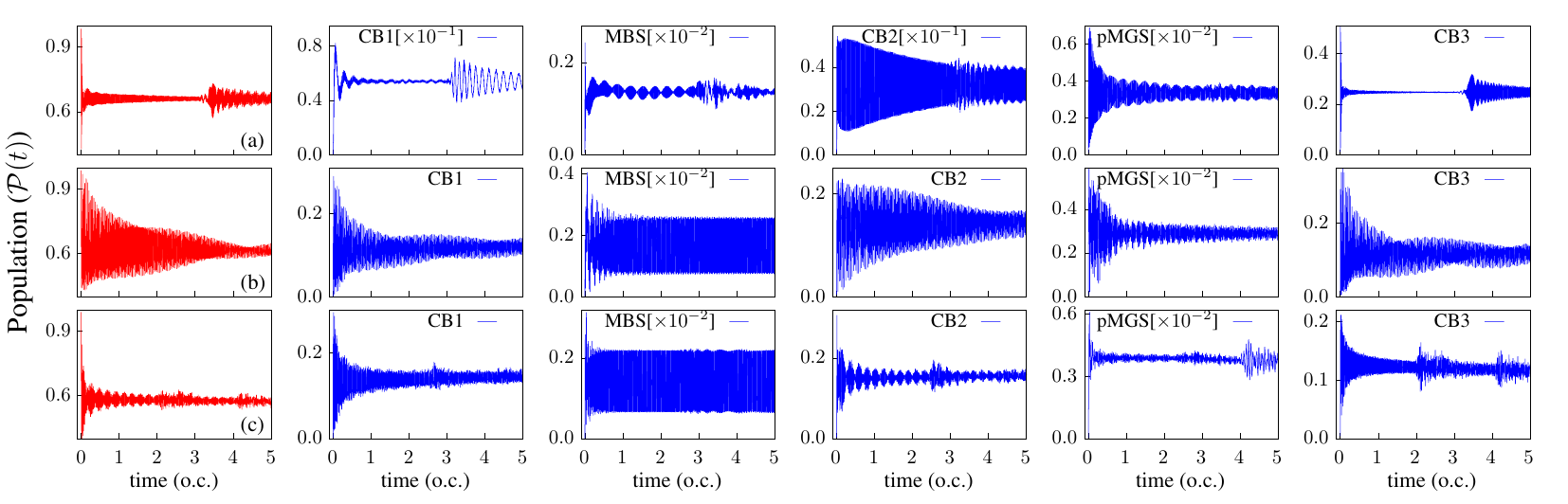}
   \caption{The figure depicts the band population of the DKC model for undriven GS. The first column in red denotes the filled GS for different tuning parameters. The rows each correspond to different values of the tuning parameter $\lambda = 0.05, 0.4, 0.8$ and depict the population of the bulk CBs and intermediate states. The other system parameters are kept similar to Fig. \ref{fig2}, expect in this case $N=300$.} 
   \label{static_pop}
\end{figure*}

Hereby, confirming the supercurrent in the static topological superconducting (TSc) systems naturally leads to an intriguing question: What would happen to the static emission profile pattern if the system were exposed to an external laser field? To proceed along this direction, we compare the results of the static GS emission profile (black line) to the driven GS harmonic spectrum to see the role of the driven laser field (pink line). We observe that the driven laser field intensifies the amplitude in addition to the plateau-like structure, which has sharp peaks in the static GS case. In Fig. \ref{fig2}(a), where $\lambda=0.05$,  at the vicinity of the orange and brown lines, individual peaks observed in the static GS case have merged and formed a plateau-like region in the driven case. The plateau is observed in the range [$\sim$ 12.8-16 eV].The reason behind this plateau-like region is the presence of multiple resonant transitions caused by the laser field. Thus, the laser field enhances the harmonic emission profile of the DKC. Earlier, in the static GS case, the dominant flow was observed due to the nMGS $\leftrightarrow$ MBS transition. The magnitude of the order remains the same due to the laser field, but the transition from VB $\leftrightarrow$ CB-III is even more enriched. This also confirms the quasi-particle nature of the TSC system, showing us that its chiral partner state is the most probable. In this state, the chiral partner state remains the most dominant transition.

In Fig. \ref{fig2}(b), we observe multiple plateaus with equiprobable intensity due to the driven laser field, where multiple peaks were seen initially in the static GS case with $\lambda=0.4$. With the driving field, the plateau is observed near the green and orange line; transitions start from  VB to the first of CB-II takes charge for the beginning of the plateau ($\sim 13.4$ eV), and the transitions from VB $\leftrightarrow$ pMGS corresponds to end of the plateau ($\sim 15.2$ eV). The static GS case shows two resonant transitions at the green and orange lines. A plateau is observed beyond the brown line around the range ($18.0-20.5$ eV), predominantly governed by VB $\leftrightarrow$ CB-III. We found a plateau around the energy range ($\sim 7.8-10.5$ eV), indicating the transitions from the last of VB $\leftrightarrow$ MBS, whereas some transitions were observed in the static case with very less magnitude around this energy range. The HHG response grows when approaching the magenta line and is followed by a peak near the red line, corresponding to transitions from VB $\leftrightarrow$ CB-I, which we did not observe earlier at $\lambda=0.05$ [refer Fig. \ref{fig2}(a)].

In the previous choice of modulating potential, the most significant plateau is seen near the orange and brown line due to quasiparticle behaviour. Similarly, in Fig. \ref{fig2}(b), the plateau is observed near the brown line, besides two other enhanced plateaus are observed with equal intensity compared to the plateau near the brown line. This is due to intermediate states (edge states) away from the bulk bands, which enhance the resonant transitions between the bands to induce multiple plateaus. We can see that the driving potential is taken to be such that it retains the quasiparticle nature of the model.

Finally, the modulation parameter is taken to be $\lambda=0.8$ [refer Fig. \ref{fig2}(c)]. The HHG spectrum forms the most significant plateau around the energy range $\sim 14-16.2$ eV, where the resonant peak is observed near the orange line in the static case. The broader bandwidth plateau begins at the magenta line ($\sim 4.5$ eV) and terminates near $\sim 9.86$ eV (VB $\leftrightarrow$ MBS). Multiple transitions between CB-I and CB-II [refer Fig. \ref{fig2}(c)] where these bulk bands are closer to each other, comparable to the other bands cause the broader bandwidth plateau. The large bandgap between VB and CB-III, leading to the less intense peak, is observed near the brown line, where no transitions are observed in the static case. Hence, the laser field enriches the intensity of the harmonics in a manner comparable to the static case, which depicts the same pattern with lower-order harmonics.
\begin{figure*}
  \includegraphics[height=6cm,width=18cm]{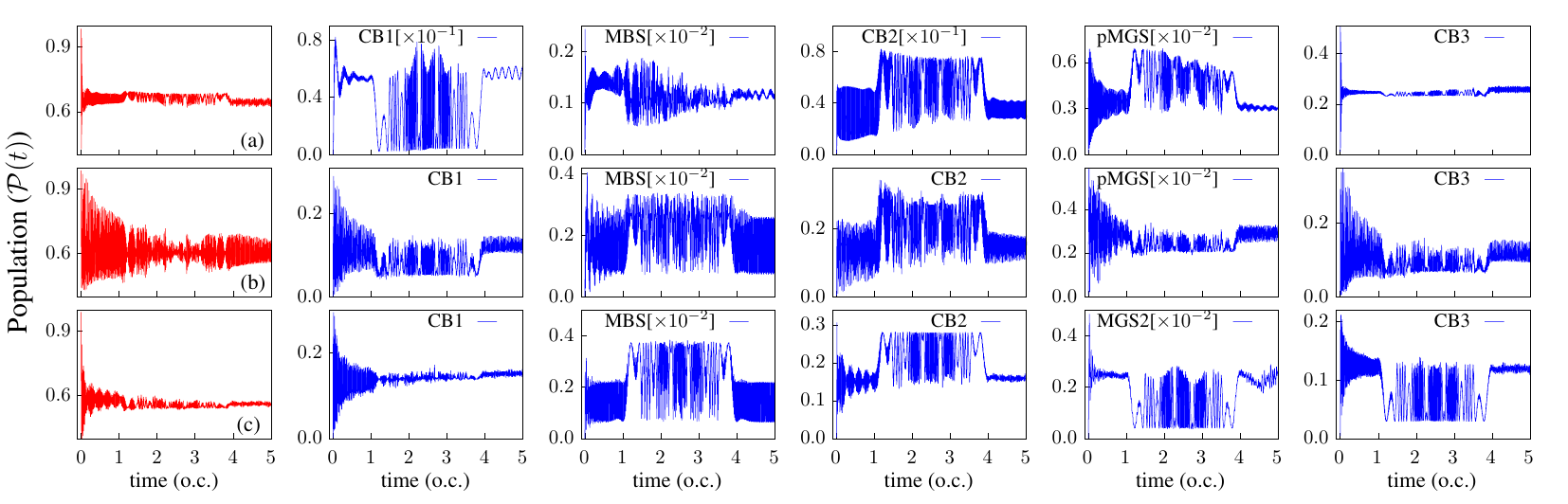}
   \caption{The figure depicts the band population of the DKC model under strong-field with the amplitude $A_0 = 0.1$ a.u. The first column in red denotes the filled GS for different tuning parameters. The rows each correspond to different values of the tuning parameter $\lambda = 0.05, 0.4, 0.8$ and depict the population of the bulk CBs and intermediate states. The other system parameters are kept similar to Fig. \ref{fig2}, expect in this case $N=300$.} 
   \label{fig3}
\end{figure*}
We now shift our attention to examining the population dynamics of the system. The study of HHG spectrum has provided several key insights: (i)The chiral partner state is the most probable state for the quasiparticle excitation; (ii) presence of the intermediate states leads to the multiple resonant excitations, and (iii) the driven laser field induces plateau-like regions with a broader bandwidth. To further clarify the dynamics of excitations in the harmonics, we analyze the band population of the system \cite{Ernotte,Nivash_23,Nivash_24}. We compute the instantaneous population of any band $m$ by defining the field-free projection operator the eigenbasis $\ket{\phi_m}$ as:
\be
\Pi_m = \ket{\phi_m} \bra{\phi_m}.
\ee
The wave function at time $t$ is expanded as $\ket{\psi(t)} = \sum_n C_n (t) e^{-iH(t)t} \ket{\phi_n}$, where the coefficient $C_n (t)= \bra{\phi_n} \psi(t) \rangle$. The population of the band $m$ is then determined by the expectation value of the field-free projection operator:
\be
\mathcal{P}_m(t) = \bra{\psi(t)} \Pi_m \ket{\psi(t)}.
\ee
We begin the investigation by analyzing the band population of the static GS, as shown in Fig. \ref{static_pop}, where the population is presented by breaking it down into the populations of the bulk conduction bands and intermediate states. The three rows of the figure correspond to different values of the tuning parameter $\lambda$. At the top row of Fig. \ref{static_pop}, where $\lambda = 0.05$, we find that the highest population occurs near the pMGS, with a probability of about $0.007$. The MBS shows a slightly lower population around $0.0024$. In this regime, the doubly-degenerate edge states exhibit higher populations compared to the bulk bands, which corresponds to the prominent peak near the orange and red lines in the harmonic spectrum, as indicated by the black line in Fig. \ref{fig2}(a). The population of CB-III is similar to that of the pMGS and higher than the MBS due to the quasiparticle behavior in this regime. In this case, CB-III has a minimum magnitude of approximately $0.51/N_b$ in the ideal case, where $N_b$ is the number of states in each band indexed by $b$.

As we increase the tuning parameter to $\lambda = 0.4$, the pMGS still has the highest population (about $0.0059$), with the MBS population rising to approximately $0.004$, as shown in Fig. \ref{static_pop}. This again matches the peak near the orange and red lines in the harmonic spectrum as shown in Fig. \ref{fig2}(b). The population of CB-III decreases to about $0.35/N_b$ (aligned with the brown line), while the populations of CB-II and CB-I are comparable  with approximate values $0.22/N_b$ and $0.28/N_b$, respectively. At this stage, the band population shows that the probabilities of all the bands' transitions are equal.

At $\lambda=0.8$, the pMGS remains the dominant state with a population of about $0.006$ (Fig. \ref{static_pop}), matching the dominant harmonic peak at the orange line in Fig. \ref{fig2}(c). The populations of CB-I and CB-II are also comparable (both around $0.3/N_b$), while the MBS population (around $0.003$) remains lower but still significant. These bulk bands (CB-I and CB-II) are nearer to the zero-energy edge states in the energy spectrum, consistent with the broadened harmonic features observed. The population of CB-III is  lower, matching with the weak peak near the brown line in Fig. \ref{fig2}(c).

Our focus on the static GS case is to confirm the quasiparticle behavior like the presence of supercurrent, and transitions preserving particle-hole symmetry in the system. This study aims to explore the population dynamics under the influence of a strong field. For this, we shift our attention to the driven GS. We present the population dynamics of the DKC model driven by the field for various tuning parameters in Fig. \ref{fig3}. First, for $\lambda = 0.05$, the most significant population is observed in the pMGS with a probability of about $0.006$. Additionally, a higher population is observed in CB-III, approximately at $0.5/N_b$. This indicates that the population of the pMGS is greater than that of CB-III, since the contribution of CB-III ($\approx 0.5/N_b$) is summed over all the states in the band ($N_b$). For the ideal case, the individual contribution of CB-III is approximately $0.0033$. Therefore, both the pMGS and CB-III have higher magnitude compared to other bands. This behavior is a result of the quasiparticle relaxation from the initially occupied VB, followed by excitation into the corresponding chiral partner state (or antiparticle state) in CB-III, as shown in Fig. \ref{fig3}(a). This behavior confirms the quasiparticle nature of the TSC system, which is consistent with the prominent plateau observed in the harmonic spectrum (Fig. \ref{fig2}(a)).

A notable population in MBS ($\approx 0.002$) is also observed in Fig. \ref{fig3}, matches with the lower intensity plateau around the red line in the harmonic spectrum shown in Fig. \ref{fig2}(b). Weak populations are observed in CB-I ($\approx 0.08/N_b$) and CB-II ($\approx 0.08/N_b$), as shown in Fig. \ref{fig3}. In the static case (refer to Fig. \ref{static_pop}), the populations of CB-I, CB-II, MBS, and MGS are lower compared to those under the influence of the driven laser field. Interestingly, the population in CB-III is similar in both scenarios, indicating the intrinsic quasiparticle nature of the system and its topological protection against the applied laser field. Previous studies have indicated that particle transitions from the initial band to higher conduction bands via a step-by-step process \cite{Jia17}, causing a time delay in the population \cite{Du17,Nivash_23}. As a result, the transition probability decreases as the particle approaches higher conduction bands, depending on the applied electric field and the system's bandgap. In our case, at the onset, the quasiparticle transitions to the highest conduction band (CB-III) with no time delay in the population. This is a key feature of the quasiparticle behavior (also known as supercurrent) in TSCs. For $\lambda = 0.4$, where multiple resonant transitions are observed in the HHG spectrum, we expect an equiprobable population transitions across all the bands. The highest population is found at the pMGS ($\approx 0.006$), with the population of CB-II ($\approx 0.35/N_b$) gradually increasing. In line with the plateau near the orange and green lines in Fig. \ref{fig2}(b), a moderate population is observed in MBS ($\approx 0.004$) in Fig. \ref{fig3}. Additionally, low populations in other bands are detected, which is consistent with the HHG spectrum. As expected from the plateau beyond the brown line in Fig. \ref{fig2}(b), a weak population is seen in CB-III ($\approx 0.3/N_b$). The enhancement near the magenta line corresponds to the population in CB-I.

Finally, for $\lambda = 0.8$, CB-I and CB-II bands are closer to each other, while the VB, MGS, and CB-III are more distanced bands (see Fig. \ref{fig2}(c)). The enhanced population in the pMGS ($\approx 0.005$) is seen in Fig. \ref{fig3}, matching the higher plateau near the orange line in the harmonic spectrum observed in Fig. \ref{fig2}(c). This results in a higher population in CB-II compared to the other bulk bands ($\approx 0.3/N_b$).  Similar populations are observed in MBS ($\approx 0.004$) and CB-II ($\approx 0.3/N_b$) as shown in Fig. \ref{fig2}(c), correlating with the large bandwidth plateau in the harmonic spectrum. Figure \ref{fig3}(c) shows the decreased population of CB-III ($\approx 0.2/N_b$), which is consistent with the lower peak near the brown line in Fig. \ref{fig2}(c).

To summarize our findings, we have shown that the static system exhibits a transition because of the intrinsic supercurrent in the TSc system. While looking at their band population, we confirm the quasiparticle nature of the model with and without the laser field. Under the influence of the laser field, enhancement in the static emission is observed, which forms a plateau-like structure. Also, these multiple transitions in the population exhibit coherent oscillatory behavior that suggests that, due to MBS, the system exhibits Rabi-like oscillations. This property is an additional takeaway of the quasiparticle nature of the system. Finally, it is important to note that our driving mechanism is a powerful tool for probing the dynamical stability of the model's topology, providing valuable insights into the system's behavior.

The authors acknowledge the Department of Science and Technology (DST) for providing computational resources through the FIST program (Project No. SR/FST/PS-1/2017/30).

\appendix
\section*{\underline{Supplementary Information}}

\section{Four-Band Problem}
\label{four}

\begin{figure}
\centering\includegraphics[width=1\columnwidth]{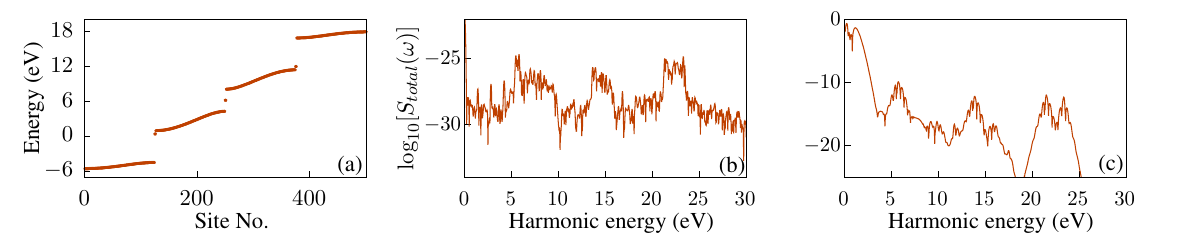}
\caption{The figure depicts the properties of a TI model, in the absence of a superconducting order parameter. The system parameters $\mu, w$ are kept similar to the main letter. Panel (a) infers to the eigen spectrum of the model shown in Eq. \eqref{4Band}. The undriven emission profile corresponding to  panel (a) in shown in panel (b). By applying laser potential, similar to the main letter, the hhg spectrum is shown in panel (c). The modualating parameter $\lambda=0.4$}
\label{Four_band}
\end{figure}

\begin{figure}[b]
\centering\includegraphics[width=1\columnwidth]{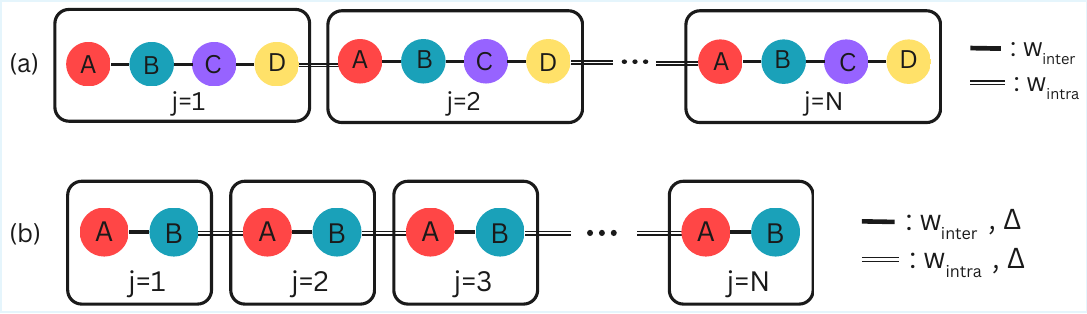}
\caption{The figure depicts the lattice structure compared in this Supplementary article. Panel (a) denotes the topological insulator (TI) model used in Sec. \ref{four} \& Eq. \eqref{4Band}, whereas panel (b) shows the topological superconductor (TSC) model discussed in Sec. \ref{delta} \& Eq. \eqref{Ham}. Four sublattices in a supercell are denoted using different colors in panel (a), where the hopping potential inside the supercell is kept as $\rm w_{inter}$ and the sublattice $D $ to $ A$ of the next supercell is taken as $\rm w_{intra}$. The TSC model has two sublattices denoted with two colors, and in addition to the previous case, there is a $p$-wave pairing term, $\Delta$, kept equal for both inter- and intra-dimer cases.}
\label{Four_band_diagram}
\end{figure}

In this section, we consider a prototype topological insulator model with four bands to elucidate the critical role of the superconducting order parameter. The Hamiltonian for the system is given by:
\begin{equation}
\begin{split}
H =  & - \sum_{j=1}^{N-1} \Bigg[  w_{\rm inter} \Big( c_{j,A}^\dagger c_{j,B} +  c_{j,B}^\dagger c_{j,C} +  c_{j,C}^\dagger c_{j,D} \Big) \\& + w_{\rm intra} \, c_{j,D}^\dagger c_{j+1,A} + \text{H.c.} \Bigg] + \mu \sum_{j=1}^{N} \sum_{\alpha} c_{j,\alpha}^\dagger c_{j,\alpha}.
\label{4Band}
\end{split}
\end{equation}
Here, $ c^\dagger $ and $ c $ are fermionic creation and annihilation operators, respectively, acting on the sublattices denoted by $\alpha \in $ $ A $, $ B $, $ C $, and $ D $ within the $ j $-th unit cell. The parameter $ w_{\rm inter} = w(1-\lambda) $ denotes the inter-dimer hopping amplitude, while $ w_{\rm intra}= w(1+\lambda) $ describes intra-dimer hopping. The chemical potential $ \mu $ is assumed uniform across all sublattices for simplicity. Figure \ref{Four_band_diagram} depicts the Hamiltonian given in the Eq. \eqref{4Band}. We set $ \lambda = 0.4 $, consistent with the parameter range used in the main paper, to obtain a similar eigen spectrum. This allows for a direct comparison and helps justify the influence of the superconducting order parameter. Figure~\ref{Four_band}(a-c) presents the eigenvalue spectrum, the undriven emission spectrum, and the driven high-harmonic generation (HHG) spectrum, respectively. Here, the initial state is kept as similar to the main letter, $\ket{\psi_{GS}}= $VB $+$ nMGS.

From Fig. \ref{Four_band}(a), we observe that the eigenvalue spectrum qualitatively resembles that of the TSC model. However, it is not symmetric about zero, which can be attributed to an energy shift by a constant factor of $6$eV. The BdG basis, used to represent TSC systems makes it symmetric to zero, by ensuring that the particle-hole symmetry in the system is preserved.  In the absence of a superconducting pairing potential, the emission spectrum shown in Fig. \ref{Four_band}(b) exhibits negligible intensity, with values on the order of $>10^{-25}$, effectively zero. This absence of emission confirms that no net current is generated in the static case without superconducting pairing, thereby emphasizing the crucial role of the superconducting order parameter in enabling both emission and the high harmonic generation (HHG) response.
\begin{figure*}
\centering\includegraphics[width=1\textwidth]{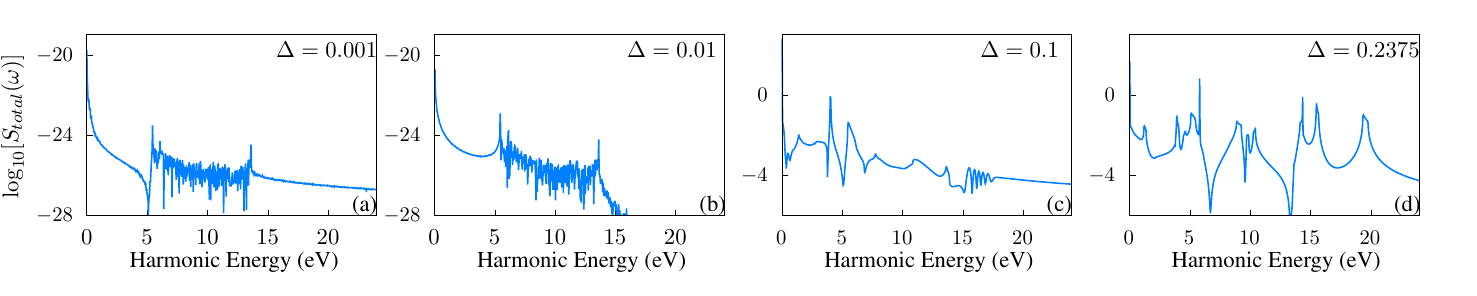}
\caption{The figure depicts the emission spectra for the static DKC model. The system parameters $\mu, w$ are kept similar to the main letter. Panel (a) infers to the case where the $p$-wave pairing potential is kept to be $0.001$ a.u. Further, we increase the $p$-wave pairing strength as (b) $0.01$ a.u.; (c) $0.1$ a.u.; (d) $0.2375$ a.u.}
\label{static_delta}
\end{figure*}
In the driven scenario, as shown in Fig. \ref{Four_band}(c), we observe a finite harmonic response, with evidence of multiple excitation and de-excitation processes in the HHG spectrum. However, the harmonic intensity monotonically decreases from the outset, lacking the plateau structure typical of strong HHG regimes.

Taken together, these results indicate that while the system without superconducting pairing can host topologically protected edge states, yet they do not contribute to a static emission in the system. The presence of a finite pairing potential is thus essential for generating observable current and strong HHG signatures in this model.


\section{Emission profile for the TSC model for various choices of \\ $p$-wave pairing potential}
\label{delta}

Having established that the system lacks an intrinsic supercurrent in the absence of a pairing potential, we now demonstrate the critical role played by the $p$-wave pairing potential in generating current. We revisit the dimerised Kitaev chain (DKC) model Hamiltonian (refer Fig. \ref{4Band}) \cite{HHG_DKC}, as introduced in the letter:

\begin{equation}
\begin{split}
H &=  \mu \sum_{j=1}^{N} \left( c_{j,A}^\dagger c_{j,A} + c_{j,B}^\dagger c_{j,B} \right) - \sum_{j=1}^{N-1} \Bigg\{ \Big[ w(1 - \lambda) c_{j,A}^\dagger c_{j,B} \\& + w(1 + \lambda) c_{j,B}^\dagger c_{j+1,A} \Big] -\Delta \Big[  c_{j,A}^\dagger c_{j,B}^\dagger + c_{j,B}^\dagger c_{j+1,A}^\dagger \Big] + \text{H.c.} \Bigg\}.
\end{split}
\label{Ham}
\end{equation}
Here, $c_{j,A}^\dagger$ and $c_{j,B}^\dagger$ are fermionic creation operators on the sublattices A and B of the $j$-th unit cell, respectively. The parameter $w$ denotes the hopping amplitude, $\mu$ is the chemical potential, and $\Delta$ is the real-valued $p$-wave superconducting pairing potential. The dimerization parameter $\lambda$ (with $|\lambda| < 1$) modulates the intra- and inter-cell hopping strengths. Figure \ref{static_delta} shows the emission spectra corresponding to the static ground state (chosen to be $\psi_{\text{GS}} = $ VB + nMGS) as a function of the pairing strength $\Delta$ in the TSC Hamiltonian of Eq. \eqref{Ham}. We observe that for $\Delta \lesssim 0.01$ a.u., no static current is generated in the system. It is important to note that the pairing potential (or superconducting order parameter) introduces effective interactions within a mean-field framework. This analysis, therefore, indicates that a finite or non-negligible pairing potential is necessary for the emergence of intrinsic supercurrents in the system.
\section{Emission profile for the DKC model for various choices of static GS}

It is important to know the role of different types of intermediate states present in the DKC model. From literature review \cite{HHG_DKC}, we name the zero energy states as Majorana bound states (MBS) and the non-zero degenerate chiral partner states ($\pm E$), which are clearly seperated from the bulk bands as mid-gap states (MGS). Depending on the sign of their energy, we classify the midgap states as $+$ve midgap states (pMGS) and $-$ve midgap states (nMGS). In this section, we focus on investigating the harmonic profile of the DKC model, by choosing these intermediate states as initial states (or GS). Like the main paper, we chose different tuning parameters and look at their harmonic profile in the static GS case.  

\noindent\underline{\textbf{nMGS as GS:}} In Fig. \ref{static_hhg}(d-f),  we begin with $\ket{\psi_{GS}}=\rm nMGS$ (blue colored line). Their corresponding eigen spectrums are presented in Fig.\ref{static_hhg}(a,b,c). We see a few peaks that correspond to different scenarios in the DKC eigen spectrum. We classified the peaks with different colored lines and listed them in the Table \ref{table}.
\begin{table}
\caption{Table classifying the color code used to characterize different emissions in the static emission. This color scheme is applicable to the main letter and the SI.}
\centering
\begin{ruledtabular}
\begin{tabular}{lcc} 
\hline
Line color & Intial state & Final state \\ \hline
Pink & Last of VB & nMGS \\ \hline
Red & MGS & MBS \\ \hline
Green & Last of VB & last of CB2 \\ \hline
Black & nMGS & pMGS \\ \hline
Orange & Last of VB & pMGS \\ \hline
Brown & Last of VB & First of CB3 \\ \hline
\end{tabular}
\end{ruledtabular}
\label{table}
\end{table}
The peaks in the emission profile, as shown in the Fig.\ref{static_hhg}(d,e,f) are responsible for the resonant transitions between the individual energy states. At $\lambda=0.05$ [refer Fig. \ref{static_hhg}(d)], we observe multiple peaks with different amplitudes (intensity of the peak). The most dominant transitions is seen around the red line ($\sim$ 6.87 eV), which is due to the MGS $\leftrightarrow$ MBS. The significant peak near the magenta line ($\sim$ 0.61 eV) is due to transitions between MGS and VB. Subsequently, a peak at the brown line ($\sim$ 14.98 eV) is felicitated due to the transitions between VB and CB3. Also, a small peak in vicinity of the orange line ($\sim$ 14.37 eV) refers to VB $\leftrightarrow$ MGS. When $\lambda$ becomes $0.4$, each band splits equidistantly, and the MGS state is finitely distanced from the bulk bands in the Fig. \ref{static_hhg}(b). This causes multiple resonant transitions that arise due to different excitations (refer Fig. \ref{static_hhg}(e)). The most significant peak near the red line ($\sim$ 5.78 eV) and a little lesser significant peak around the sea-green line ($\sim$ 14.2 eV) corresponds to the transition between VB and the last CB2, which is not observed in the previous choice of $\lambda$. A peak is observed near the vicinity of the orange line ($\sim$ 15.45 eV) followed by a less significant peak at the brown line ($\sim$ 19.32 eV). Finally, the smallest peak is observed around the magenta line ($\sim$ 3.87 eV) which infers to the transitions between VB and MGS. Moreover, for $\lambda=0.8$ [refer \ref{static_hhg}(c, f)], we see the most significant peak near the orange line ($\sim$ 15.22 eV). Additionally the similar peaks are seen around the red line ($\sim$ 5.36 eV) followed by multiple broad peaks (plateau-like structure) up to 10 eV due to the band closing between CB1 and CB2 near zero energy states. There is no peak obtained at the magenta line ($\sim$ 4.5 eV) and brown line ($\sim$ 19.72 eV), but after the brown line ($\sim$ 22.4 eV), tiny oscillations appear due to transitions from VB to last CB3.

\noindent \underline{\textbf{MBS as GS:}} On the other hand, if we look at the resonant excitations when $\ket{\psi_{GS}}=\rm MBS$ for the similar modulating parameters like the previous case, as shown in Fig.\ref{static_hhg}(g-i), we are able to see different behavior in the emission profile. In the Fig. \ref{static_hhg}(g), the $\lambda$ takes the value $0.05$, in which the transitions between nMGS and pMGS fecilitate the most significant peak is observed around the black line ($\sim$ 13.75 eV). Additionally, no transitions are observed near the orange and brown lines (which corresponds to MGS $\leftrightarrow$ CB3, and VB $\leftrightarrow$ CB3 respectively). A small peak around the magenta line ($\sim$ 0.61 eV) corresponds to transitions from VB to nMGS. Figure \ref{static_hhg}(h) reveals the current when $\lambda =0.4$ Here, the most significant peak is obtained near the black line ($\sim$ 11.57 eV) that corresponds to the transition between two chiral MGS partners (i.e. pMGS $\leftrightarrow$ nMGS). When $\lambda = 0.4$, as mentioned earlier, the harmonics are enriched due to the non-triviality in the model. Hence the resulting harmonics observed are near the orange line ($\sim$ 15.45 eV) and the magenta line ($\sim$ 3.87 eV), followed by a less intense peak at the brown line ($\sim$ 19.32 eV). Finally, looking at the system when $\lambda=0.8$ [Fig.\ref{static_hhg}(i)], the most significant peak is obtained around the black line ($\sim$ 4.5 eV) and a lesser significant peak near the orange line ($\sim$ 15.22 eV). Additionally, a peak at the magenta line ($\sim$ 4.5 eV) is followed by tiny oscillations observed after the brown line ($\sim$ 22.4 eV). The motivation to analyse with the intermediate states being our choice of GS is to reveal the non-triviality in the model. We further discussed the more realistic or the conventional choice of GS in the main letter.

\noindent \underline{\textbf{VB as GS: }} To investigate the role of bulk states in the GS, we computed the emission profile when the GS is summed over the VB. The emission profile is shown in the Fig. \ref{static_hhg}. We see that there is no sufficient emission in the profile. This investigation directly confirms that the GS degeneracy is the reason for the presence of emission in the static case. When the degeneracy is excluded, the model doesnot have emission.
\section{Population dynamics for the static GS}

In this section, to investigate the band population in the static DKC \cite{Ernotte, Nivash_23, Nivash_24}, we chose the ground state to be either the MGS or the MBS, denoted by $\ket{\psi_{GS}}=$ MGS and MBS. All system parameters are kept similar to the emission profile analyses. 

\begin{figure}
\centering\includegraphics[width=1\columnwidth]{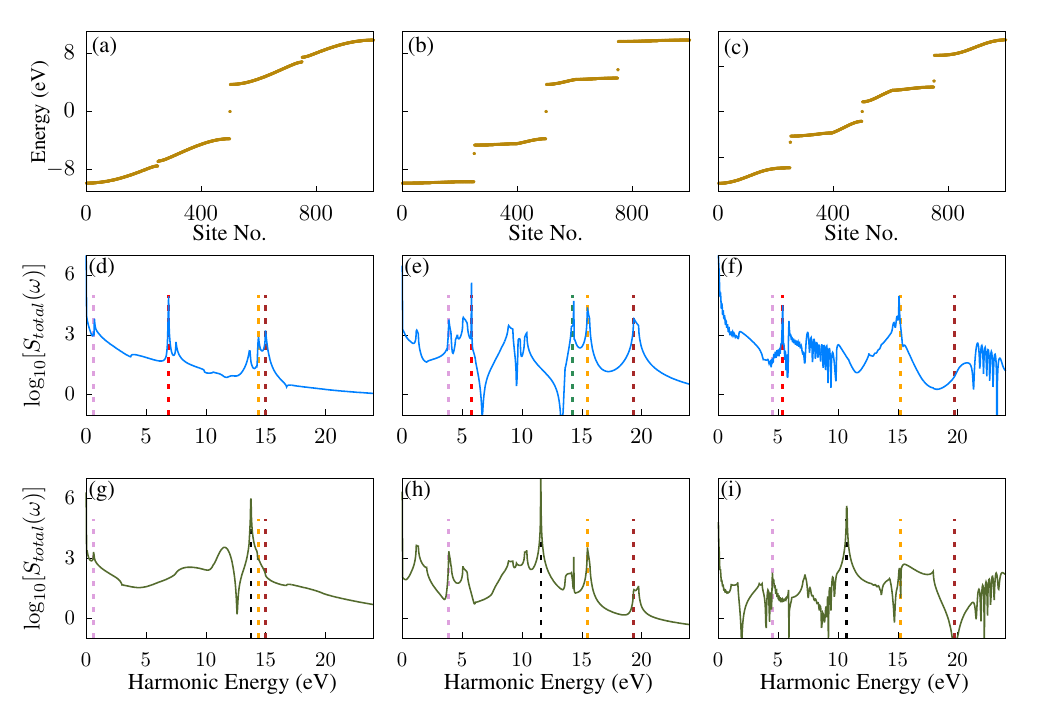}
\caption{The figure reveals the eigen spectrum and emission profile for different GS with three different modulating parameters ($\lambda=0.05, 0.4, 0.8 $). Panels (d-f) and (g-i) denote the emission profile for nMGS and MBS as GS respectively.}
\label{static_hhg}
\end{figure}
\begin{figure}[b]
  \includegraphics[width=1\columnwidth]{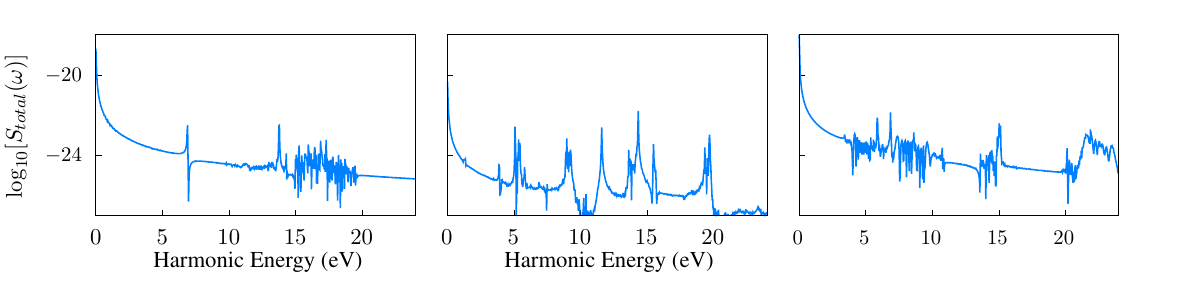}
   \caption{The figure reveals the emission profile for GS to be summed over all the states in the VB, with three different modulating parameters, $\lambda=0.05, 0.4, 0.8 $.} 
   \end{figure}

\begin{figure*}
  \includegraphics[width=1\textwidth]{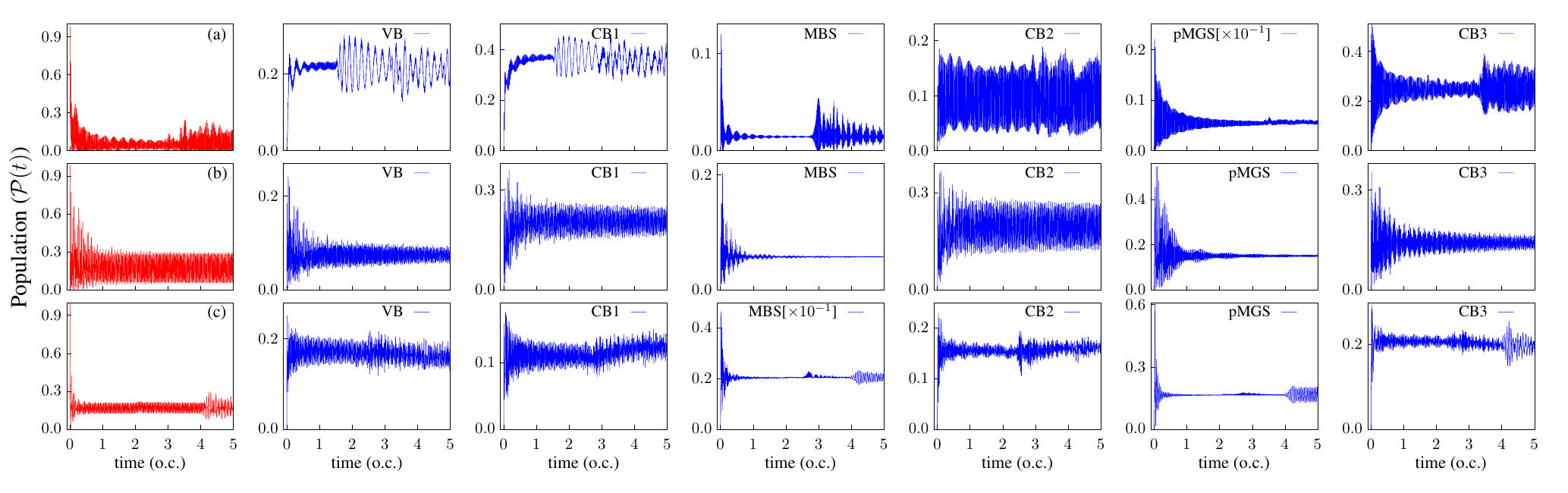}
   \caption{The figure displays the band poulation for the static GS to be $\ket{\psi_{GS}}= \rm MGS$. The rows corresponds to different values of $\lambda$: $(\rm a) 0.05, (\rm b) 0.4, {\rm and} (\rm c) 0.8$.} 
   \label{pop_mgs}
\end{figure*}

\noindent \underline{\textbf{nMGS as GS:}}  We begin our study with the choice, $\ket{\psi_{GS}}=\rm nMGS$, and the corresponding band population results are presented in Fig.\ref{pop_mgs} for different modulating parameters $\lambda=$ 0.05, 0.4, 0.8. This figure clearly demonstrates how the band is populated by differentiating the bulk bands from intermediate states. The first row denotes the bands corresponding to $\lambda = 0.05$, second for $\lambda =0.4$ and third to $\lambda=0.8$. Beginning with $\lambda=0.05$, the most significant population is observed in MBS ($\approx$ 0.12) [Fig.\ref{pop_mgs}(a)], which is corroborated by a most dominant peak at the red line in the emission profile (presented in the Fig. \ref{static_hhg}(d)). Also the next prominent population is seen in pMGS ($\approx$ 0.02), which coincides with peak near the orange line in the emission profile (presented in the Fig. \ref{static_hhg}(d)). We already observed that the MGS is in close proximity to the bulk band CB1 than VB, which makes it more viable for a quasiparticle to populate CB1 than VB. This results in the higher population amplitude in CB1 ($\approx$ 0.4) and lower magnitude in VB ($\approx$ 0.2). One should understand that the population in the bulk bands is summed over all states, i.e., N=148 represents the total number of states in each bulk band, whereas the intermediate states are summed over doubly-degenerate states. So, for an ideal case the range will be $\approx$ 0.4/N for CB1 and $\approx$ 0.2/N for VB are shown in Fig.\ref{pop_mgs}(a) row. The bulk band population is smaller than the intermediate states (MBS/MGS) population, which is distributed over doubly-degenerate states. Furthermore, a high population is observed in CB3 ($\approx$ 0.45/N) and less significant population in CB2 ($\approx$ 0.2/N) [Fig.\ref{pop_mgs}(a)]. It is correlated with the peak near the vicinity of the brown line in the emission profile (presented in the Fig. \ref{static_hhg}(d)). The highest population among the bulk bands is to CB3. This is due to the quasi-particle behavior in the model.
\begin{figure*}
  \includegraphics[width=1\textwidth]{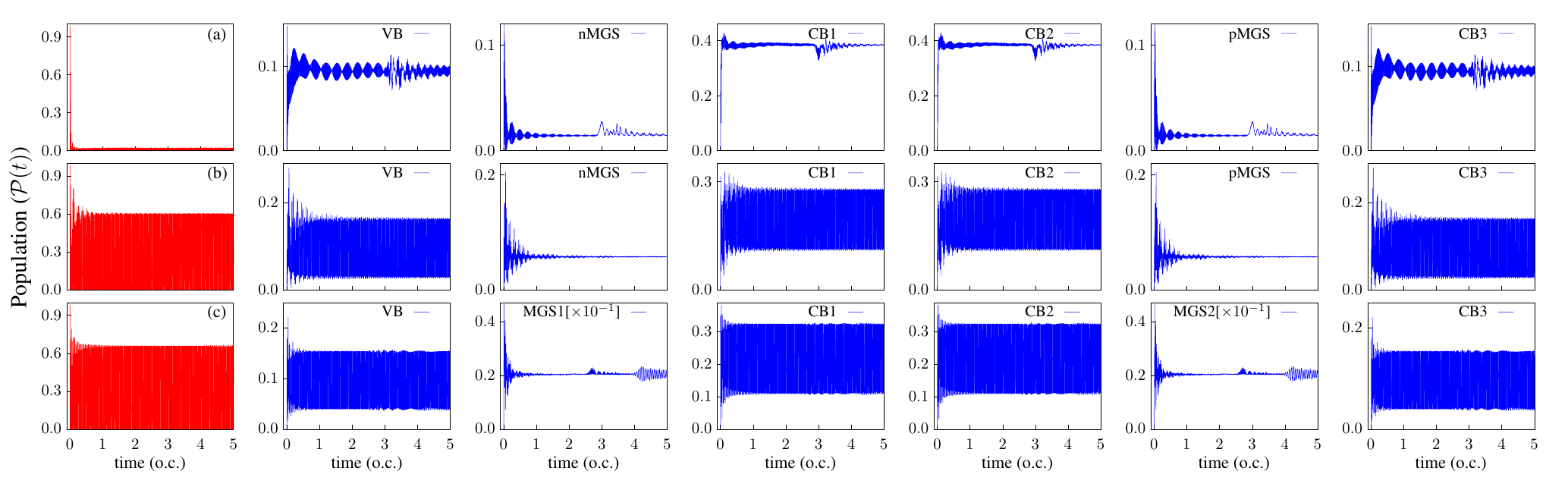}
   \caption{The figure displays the band poulation for the static GS to be $\ket{\psi_{GS}}= \rm MBS$. The rows corresponds to different values of $\lambda$: $(\rm a) 0.05, (\rm b) 0.4, {\rm and} (\rm c) 0.8$.} 
   \label{pop_mbs}
\end{figure*}
When $\lambda=0.4$, the pMGS ($\approx$ 0.5) has the most significant population, followed by the MBS ($\approx$ 0.2), as seen in Fig.\ref{pop_mgs}(b). This explains the emission intensity around the orange and red lines in the emission profile (presented in the Fig. \ref{static_hhg}(e)). Compared to the case of $\lambda=0.05$, the quasiparticle population in CB1 and VB is considerably reduced, which can be attributed to the MGS isolated from the bulk bands. As a result, the population in CB1 remains relatively higher ($\approx$ 0.3/N), while the population in VB is significantly lower ($\approx$ 0.22/N), as shown in Fig.\ref{pop_mgs}(b). The significant population is seen in both the bulk bands CB3 ($\approx$ 0.35/N) and CB2 ($\approx$ 0.35/N) as shown in Fig.\ref{pop_mgs}(b), which is corroborated with the peak near the vicinity of the brown and green line in the emission profile (presented in the Fig. \ref{static_hhg}(e)). This equiprobable population arises from equi-probable band splitting, which leads to multiple transitions. For $\lambda=0.8$, pMGS($\approx$ 0.6) hosts the dominant population in Fig.\ref{pop_mgs}(c), which results in the most significant population that corresponds to the peak at the orange line in the emission profile (presented in the Fig. \ref{static_hhg}(f)). A smaller population is seen in MBS ($\approx$ 0.04), which leads to a significant peak near the red line in the emission profile (presented in the Fig. \ref{static_hhg}(f)). In this case, MGS is away from the bulk bands, resulting in less quasiparticle excitation in CB1 and VB than in the previous modulating potential. The higher population magnitude in VB ($\approx$ 0.21/N) and lesser magnitude in CB1 ($\approx$ 0.12/N). Due to the bulk band, CB1 and CB2 are closer to the MBS, which significantly enhances the population in CB2 ($\approx$ 0.22/N), giving rise to a series of broad peaks extending up to 10 eV in the emission profile (presented in the Fig. \ref{static_hhg}(f)). Furthermore, a significant population is found in the CB3 ($\approx$ 0.3/N) [refer Fig.\ref{pop_mgs}(c)], which correlates with the small oscillations appearing just beyond the brown line in the emission profile (presented in the Fig. \ref{static_hhg}(f)).

\noindent \underline{\textbf{MBS as GS:}}  Now, we look at the static band population when $\ket{\psi_{GS}}=\rm MBS$ for a similar modulating parameter like the previous case, as shown in Fig.\ref{pop_mbs}. When $\lambda=0.05$, both the MGSs have a similar magnitude and hosts the most dominant population in Fig.\ref{pop_mbs}(a), which corroborates with the most significant peak at the red line in the emission profile (presented in the Fig. \ref{static_hhg}(g)). The MBS is located at the midpoint of the chain. This causes symmetric population distributions across bands with respect to the MBS, like pMGS $\&$ nMGS, CB1 $\&$ CB2, and VB $\&$ CB3. This yet again confirms that the most probable state is the chiral partner state, which is a signature of the quasi-particle behavior of the model. Beyond MGS chiral partners states, the population in CB1 $\&$ CB2 is slightly high ($\approx 0.1/N_b$), followed by even lesser population in VB $\&$ CB3 are shown in Fig. \ref{pop_mbs}(a). In the case of $\lambda=0.4$, a highest population magnitude is observed in the pMGS $\&$ nMGS [Fig.\ref{pop_mbs}(b)], corroborating with the most dominant peak at the red line in the emission profile (presented in the Fig. \ref{static_hhg}(h)). Followed by a lesser population in CB1 $\&$ CB2 and even lower magnitudes in VB $\&$ CB3. The CB3 population correlated with the small peak near the brown line in the emission profile (presented in the Fig. \ref{static_hhg}(h)). For $\lambda=0.8$, both the MGSs have the dominant population, which is reflected with a peak near the red line in the emission profile (presented in the Fig. \ref{static_hhg}(i)). Fig.\ref{pop_mbs}(c) shows the significant population in the CB1 $\&$ CB2 and a less significant population in the VB $\&$ CB3.

To conclude, we show that by including the intermediate state as a initial state, the emission profile is enriched with coherent transitions that are explicity visible in the band population. Since, for choice of VB as GS did not give emission, we did not discuss its population here.

\end{document}